
\def\lsim{ {\raise -3mm \hbox{$<$} \atop \raise 2mm
\hbox{$\sim$}} }
\def\gsim{ {\raise -3mm \hbox{$>$} \atop \raise 2mm
\hbox{$\sim$}} }
\def\km{{\rm km}}
\def\cm{{\rm cm}}
\def\sec{{\rm s}}
\def\day{{\rm day}}
\def\gram{{\rm g}}

\def\kT{{\rm kT}}

\def\GeV{{\rm GeV}}
\def\svec{{\vec S}}
\def\selectron{{\tilde e}}

\def\photino{{\tilde\gamma}}

\baselineskip = 21pt
\centerline{\bf A New Technique for Detecting Supersymmetric Dark
Matter}
\bigskip
\centerline{Glenn D. Starkman}
\centerline{Canadian Institute for Advanced Research, C.I.T.A.,
60 St. George St., Univerity of Toronto,}
\centerline{Ontario M5S 1A7, Canada}
\centerline{\& David N. Spergel}
\centerline{Princeton University Observatory, Princeton, NJ 08544}
\bigskip

{\bf
     We estimate the event rate for excitation of atomic transition
by
photino-like dark matter.  For excitations of several eV, this
event rate can exceed naive cross-section by many orders of
magnitude.
Although the event rate for these atomic excitation is smaller than
that of nuclear recoil off of non-zero spin nuclei, the photons
emitted by the deexcitation are easier to detect than low-energy
nuclear recoils.  For many elements, there are several low-lying
states with comparable excitation rates, thus, spectral ratios
could be used to distinguish signal from background.
}

     The composition of most of the mass in our Galaxy is not
known.
Dynamical estimates of the mass of the Galaxy suggests that it
exceeds the mass in luminous stars, gas and dust
by an order of magnitude$^{1}$.
This problem is not unique to our Galaxy, but appears ubiquitous:
as dynamical estimates of the density of the universe$^2$ suggest that
it  is also an order of magnitude larger than
the baryon density inferred from light element abundances
and big bang nucleosynthesis$^3$.
Thus, most of our Galaxy may be composed not of ordinary baryonic
matter but of some exotic non-baryonic component.

     Supersymmetry, an elegant and well-studied extension
of the standard  model of particle physics, suggests a particle physics
solution
to this astronomical quandary.
In the minimal supersymmetric model, the neutralino,
the lightest mass eigenstate that carries a new conserved charge,
R parity, may be stable.  This particle,
a linear combination of the photino, higgsino and zino interaction
eigenstates,
would have been created in the early universe  and can be stable.
For a wide range
of parameters  in the minimal supersymmetric model, its density
would exceed the density in baryons$^4$.

     In the past few years, a series of experiments at LEP and
Fermilab appear to hint that supersymmetry may be real and that the
supersymmetry scale may be just beyond that probed by the current generation
of accelerators.  In grand unified theories (GUTs) based solely on
the standard model,  $\sin^2 \theta_W$, the electroweak mixing angle, is
predicted$^{5,6}$ to be  $0.215\pm0.002$,
while in SUSY GUTs, it is predicted$^{7,8}$ to be $0.233\pm0.002$.  These
predictions were made at a time in which this angle was known to lie between
$0.2$ and $0.25$.  Experiments at LEP find$^9$
 $\sin^2 \theta_W  = 0.233\pm0.001$,
10 $\sigma$ away from the standard model prediction.  In SUSY GUTs, the top
mass is predicted$^{10,11}$ to be between 170 and 190 GeV, while GUTs based on
the standard model predict a top mass around 220 GeV.   Recent
experiments (CDF Collaboration, Fermilab-PUB-94/097-E)
at the Fermilab Tevatron
imply a top quark mass of $174 \pm 15$.
These experimental hints strengthen the motivation to search for
supersymmetric dark matter.

     If the neutralino did indeed compose most of the mass of the
Galaxy, then the flux of \break
$(10^7\GeV/m_x) {\rm particles}/\cm^2/\sec$ is potentially
detectable. (Here, $m_x$ is the neutralino mass.)
The electron interacts primarily with the photino component
of the neutralino via the exchange of scalar
selectrons, $\tilde e$.
Neutralinos interact with quarks predominately through the exchange of
scalar quarks, $\tilde q$ and scalar
higgsinos, $\tilde h$.   In the non-relativistic limit, appropriate
for dark matter particles moving in our galaxy at $v \sim 300$
km/s, the scattering amplitude for the interaction with electrons,
$\tilde M \simeq {4 e^2 \over M_{\tilde e}^2} \vec S_\photino \cdot \vec S_e$,
where $S_\photino$ is the photino spin, $S_e$ is the electron spin,
$e$ is the electron charge and $M_{\tilde e}$ is the selectron mass.
The scattering amplitude
for quarks$^{12}$ is similar with the electron charge, spin and selectron
mass replaced by the quark charge, spin and squark mass.
Unlike squark exchange, the higgsino exchange is spin independent
and thus is important is scattering off of low spin nuclei.

     Searches for neutralinos have attempted to detect their
scattering elastically off of nuclei through squark exchange.  This elastic
scattering event deposits several keV of kinetic energy in the nucleus.  In a
solid state detector, this kinetic energy is converted into electron-hole pairs
and phonons, which are potentially detectable in a sensitive low
background detector.   Because of the low event rate and the tiny energy
deposition, this is a very challenging experiment.  The difficulty of this
experiment is further compounded by the limited range of materials that can
be used: neutralino scattering cross-sections are suppressed in
materials with  an even number of protons.
Despite heroic effort, the experimental limits on weakly
interacting dark matter have not significantly improved in the decade since
these experiments were proposed and are still several orders of magnitude
away from the parameter range relevant for supersymmetric dark
matter.

     In this letter, we explore an alternative
approach to detecting neutralinos:  detecting their inelastically
exciting atomic states.  This possibility
has not been carefully explored in the past as a naive estimate
of the cross-section would suggest that it is suppressed relative
to nuclear scattering by the square of the ratio of $m_e$,
the mass of the electron, to
$m_n$, the mass of the nucleus.  As we will
show in this letter,
there are several mitigating factors that balance
this $4 \times 10^6 A^2$ suppression factor:
(1) there is a kinematic enhancement of
${\cal O}\bigl( (\Delta E/v_\photino)/(m_e v_\photino))^2\bigr)
\simeq {\cal O}\bigl(10^{-2}(\alpha_{fs}c/v)^4\bigr) \simeq 10^2$
(the $10^{-2}$ being an approximate numerical factor due to the higher
overlap integrals for smaller $\Delta E$'s);
(2) the electron charge is either (3/2) or 3 times
larger than the
quark charge, leading to an enhancement of
up to $3^4$; (3) the selectron mass is expected to be
significantly smaller that the squark mass,
this enhances the atomic scattering processes
by another factor of perhaps $2^4$; (4) the possibility of exciting in a
given atom any of the outer shell electrons to one of several
possible states can lead to an enhancement in the overall event
rate by another factor of a few.   These four factors, when
combined, enhance the cross-section by a factor of few $\times10^5$
and makes this process worthy of more detailed study.

Atomic excitation may also have several experimental advantages
over nuclear scattering.  As several different energy levels are
excited at predictable rates, neutralino should have a clean
experimental signature.  In nuclear recoil interactions,
neutralinos couple only to nuclei with non-zero spin, a relatively
limited set of materials can be used in these experiments.  A much
broader class of materials can be used in an atomic excitation
experiment and a broader class of detection techniques are
available for detecting several eV photons.

As the electron-photino interaction can be approximated
as point-like, the cross-section for exciting an atom from
from an initial state with wave function $\psi_1$ and energy $E_1$
to a final state with wave function $\psi_2$ and energy $E_2$,
the cross section is easily calculated:
$$\eqalignno{\sigma v = &{4 e^4\over m_\selectron^4}
\vert< f~\vert \svec_e \vert ~i>\vert^2
{1\over (2\pi)^6} \int{d^3 q^\prime_\photino\over (2\pi)^3}
                {d^3 q^\prime_N       \over (2\pi)^3}
\biggl\vert\int d^3r d^3r_N d^3r_\photino
         \psi_1({\vec r})\psi_2^\star({\vec r})
     e^{i({\vec q}_\photino -{\vec q}_\photino^\prime)\cdot{\vec
r_\photino}}
        e^{-i{\vec q}_N(1 + m_e/m_N)\cdot{\vec r}_N}
\times \cr \times &
     \delta^{(3)}({\vec r}_\photino - {\vec r}_e)\biggr\vert^2
     2\pi \delta(E_\photino - E_\photino^\prime - (E_2-E_1))\  .
&(1)\cr}$$
Here  ${\vec r} = {\vec r}-{\vec r_N}$, (un)primed quantities refer
to the (initial) final state,
and we have neglected the motion of the nucleus
relative to the atomic center of mass.
We have also approximated the electron-photino interaction as
pointlike, which
is valid for $m_\selectron >> E$.

Using $ \delta^{(3)}({\vec r}_\photino - {\vec r}_e)$, we eliminate
the integral over
$d^3r_\photino$.  The integral $d^3r_N$ gives a delta function of
3-momentum;  which is used to eliminate the $d^3q_\photino^\prime$
integral.
The remaining energy delta function is used to do the integral
$d\cos\theta_N$
over the angle between the   outgoing nuclear recoil momentum and
the
incoming photino momentum.  We are left finally with
$$\sigma v \simeq {1\over 2 \pi} {4 e^4 \over m_\photino^4}
\vert< f\vert \svec_e \vert i>\vert^2
{1\over \beta_{\photino}}
\int_{q_{min}}^{q_{\max}}q_Ndq_N
\biggl\vert\int d^3r\psi_1({\vec r)}\psi_2^{\star}({\vec r)}
e^{i{\vec q}_N\cdot {\vec r}}\biggr\vert^2 \   . \eqno(2)$$
$\beta_\photino$ is the photino velocity divided by $c$;
$q_{min} \simeq (E_2-E_1)/v_\photino$ and $q_{\max}\simeq 2 m_N
\beta_\photino$
are the maximum and minimum nuclear recoil momenta allowed by
kinematics,
obtained by enforcing $\vert\cos\theta_N\vert\leq 1$.

Assuming that a Gaussian distribution for the dark
matter particles impinging on the Earth, $f(v) = \exp[(\vec v -
\vec v_e)^2
/2 \Sigma^2]$ yields a average interaction rate:
$$\eqalignno{<\sigma v> &\simeq
      {2\over \pi} \vert<f\vert\svec\vert i>\vert^2
          {16\pi^2\alpha^2\over m_\selectron^4 a_0^2}
          { c\over \Sigma{\sqrt 2}} g_{ij} {\cal J}&(3)\cr
&\simeq 2.75\times 10^{-33}\ {{\rm events}\over {\rm atom}-\day}
 {\cal J} g_{ij} {\vert<f\vert\svec\vert i>\vert^2\over 3/4}
\biggl({m_x \over 100 \GeV}\biggr)
\biggl({100\GeV\over m_\selectron}\biggr)^4
\biggl({\rho_{DM} \over .4\GeV\cm^{-3}}\biggr)
\biggl({250\km/\sec\over \Sigma{\sqrt 2}}\biggr)\cr
&\simeq {1.7\  {\rm events}\over \kT-\ \day} {\cal J} g_{ij}
	\biggl({m_x \over 100 \GeV}\biggr)
     \biggl({100\GeV\over m_\selectron}\biggr)^4
	\biggl({\rho_{DM} \over .4\GeV\cm^{-3}}\biggr)
	 \biggl({\rho_{detector} \over A \
     \gram \cm^{-3}}\biggr)
\biggl({250\km/\sec\over \Sigma{\sqrt 2}}\biggr)}  $$
$g_{ij}$ is  the product of the (average) occupation number of the
initial state, $o_i$  ($0\leq o_i \leq 1$) and $1-o_f$.
All of the atomic structure is contained in the factor,
${\cal J} \leq O(1)$, and can be
can be calculated numerically from
$$\eqalign{{\cal J} &=    {\sigma \over \sqrt{2} v_e}
          \int_0^\infty {\ a_0^2 dq\ q} \biggl\vert\int d^3r
            \psi_1({\vec r})\psi_2^\star({\vec r})
            e^{i{\vec q}\cdot{\vec r}} \biggr\vert^2 \times
            \cr
&\times\left\{ {\rm erfc}\left[ \max \left({\Delta E\over q\sigma{\sqrt 2}},
                   {q\over\sigma{\sqrt 2}m_N\sigma}\right)-
               {v_e\over \sigma{\sqrt 2}}\right]
            -{\rm erfc}\left[ \max \left({\Delta E\over q\sigma{\sqrt 2}},
                  {q\over\sigma{\sqrt 2}m_n\sigma}\right)+
               {v_e\over \sigma{\sqrt 2}}\right]\right\}\cr &} \eqno{(4)}$$
using numerical wave functions for any atom (or molecule) of
interest.
We have calculated
${\cal J}$ for a variety of
transitions in selected atoms.  For energy levels separated by
a few eV, ${\cal J}$ is between 0.1 and
1 for most atomic states.   When these excited states decay emitting few eV
photons, these photons may be more easily detectable  than the
phonons and electron-hole pairs produced through nuclear
scattering.

Any experiment that seeks to detect the rare photons
produced by neutralino excitation of atoms faces several
challenges.  The detector will have to be very massive, very quiet,
and placed in a low-background environment such as the Gran Sasso
tunnel.   It will have to be cooled to a temperature
less than $135 (E/$eV)$^{-1}$ to avoid thermal photon background.
The atom will have to deexcite predominantly by spontaneous decay
(with the emission of a photon) and not by collisional de-excitation.
The material will have to be transparent enough so that the
photon is not degraded through reabsorbtion followed by
collisional deexcitation.
This concern may be addressed by focusing on those transitions
which frequently decay
to the ground state by a cascade of two or more photons,
rather than directly.  In this case only
the last of these photons would resonantly re-excite  the ground state;
while the others would scatter mostly elastically.

There are, on the other hand, several sources of encouragement
for the experiment.  In many materials, the
event rates are interesting for excitation to excited states.
Thus, there are potentially several different lines that
may be detectable.  The event rate in each of the lines
will be annually modulated by the motion of the Earth around the
Sun with an amplitude that varies in a predictable way from
line-to-line.  In many atoms and molecules, there are a variety of
fine structure splitting states and isoelectronic states
separated from the ground state by a few eV.  Thus, there
is a wide variety of materials that can be used in the detector.
The authors will provide to interested experimentalists
(or collaborate with them in the calculation of)
event rates in various materials.

While the current generation
of scintillator detectors designed for solar neutrino work
do not appear to be sensitive to the eV photons discussed here,
there are a wide variety of detectors that can be used.
In some materials, in which the electron orbitals are aligned along
a particular direction, such as liquid crystals, there
is the possibility of a directional dependence in the signal.

The  challenge to experimental physicists,
material scientists and chemists is to identify low noise
eV detectors for large volume targets.
The potential reward is the ability to detect the material
that composes most of the mass of the Galaxy and find the first
Supersymmetric particle.

Acknowledgements:
The authors thank F. Calaprice for useful discussions.
GDS would also like to thank
B. Cabrera, D. Caldwell,  D. Cline,
K. Griest, M. Kamionkowski and A. McDonald
for useful interactions; M. Schmidt for supplying and assistance with
the GAMESS$^{13}$ program; W. Keogh and A. Boothroyd for their
advice in matters of computational chemistry;  C. Bunge
for discussions and proferred assistance in calculating
excited state wave functions. DNS acknowledges the NSF
and NASA for support and thanks CITA for its hospitality and
GDS acknowledges NSERC for suppport and
thanks Princeton University Observatory  for its hospitality.

References:

\noindent$^1$ Fich, M. and Tremaine, S., Ann. Rev. Astron. Astrophys.
{\bf 29}, 409 -- 445 (1991)

\noindent$^2$ Dekel, A.,  Ann. Rev. Astron. Astrophys., in press.

\noindent$^3$ Walker, T., et al. 1991, ApJ, {\bf 376}, 51 -- 69 (1991).

\noindent$^4$ Ellis, J. et al. 1984, Nucl. Phys. B, {\bf 238}, 453 - 480
(1984).

\noindent$^5$ Georgi, H. and Glashow, S.,  Phys. Rev. Lett {\bf 32}, 438-441
(1974)

\noindent$^6$ Georgi, H., Quinn, H. and Weinberg, S.,  Phys. Rev. Lett {\bf
33}, 451-454 (1974).

\noindent$^7$ Dimopoulos, S., Raby, S. and Wilczek, F.,  Phys. Rev. {\bf D24},
1681-1683 (1981).

\noindent$^8$ Dimopoulos, S. and Georgi, H.,  Nucl. Phys. {\bf B193}, 150-162
(1981).

\noindent$^9$  Langacker, P. in "Review of Particle Properties," Phys. Rev.
{\bf D45}, III.59-64
(1992), and references therein.

\noindent$^{10}$  Ananthanarayan, B., Lazarides, G. and Shafi, Q.,  Phys. Rev.
{\bf D44}, 1613-1615 (1991).

\noindent$^{11}$ Arason, H., Castano, D., Keszthelyi, B.,
Mikaelian, S., Piard, E.J., Ramond, P. and Wright, B.,
Phys. Rev. Lett. {\bf 67}, 2933-2937 (1991).

\noindent$^{12}$ Goodman, M. and Witten, E.,  Phys. Rev. {\bf D31}, 3059-3063
(1985).

\noindent$^{13}$ Schmidt,M.W., Baldridge,K.K., Boatz, J.A., Elbert, S.T.,
              Gordon, M.S., Jensen, J.H., Koseki, S., Matsunaga, N.,
          Nguyen, K.A., Su, S.J., Windus, T.L., Dupuis, M., Montgomery, J.A.,
                   J.Comput.Chem.  14, 1347-1363 (1993)
\par \vfill \end